\documentclass[twocolumn,aps,showpacs]{revtex4}
\usepackage{graphicx}

\begin{document}

\title{Vortex solitons in an off-resonant Raman medium}

\author{A.V. Gorbach and D.V. Skryabin}
\affiliation{Centre for Photonics and Photonic Materials, Department
of Physics, University of Bath, Bath BA2 7AY, UK}

\author{C.N. Harvey}
\affiliation{School of Mathematics and Statistics, University of Plymouth, Plymouth
PL4 8AA, UK}

\begin{abstract}
We investigate existence and linear stability  of coupled vortex
solitons supported by cascaded four-wave mixing  in a Raman active
medium excited away from the resonance.  We present a detailed
analysis for the two- and three-component vortex solitons and
demonstrate the formation of stable and unstable vortex solitons,
and associated spatio-temporal helical beams, under the conditions
of the simultaneous frequency and vortex comb generation.
\end{abstract}

\maketitle

\section{Introduction}
Optical vortices are point phase singularities of the
electromagnetic field, with  the beam intensity vanishing  at the
singularity and the field phase changing by $2\pi l$  along any
closed loop around it.  $l=0,\pm 1,\pm 2,\dots$ is known as the
orbital angular momentum quantum number or vortex charge. In a
nonlinear medium vortices  can propagate undistorted due to a
balance between diffraction and nonlinearity, and form so-called
vortex solitons \cite{Desyatnikov}. Nonlinearity  can also trigger
frequency conversion accompanied by the conversion of the charge
$l$. In particular, in the second harmonic generation process, the
fundamental field carrying a vortex with the charge $l$ is converted
into the second harmonic field with the charge $2l$
\cite{Desyatnikov,Buryak2002,DSP+1996,SF1998,Torres1998a,DCM+2000}.
Analogous conversion rules have been reported for the degenerate
four-wave mixing in Kerr-like materials \cite{Mihalache2003} and for
the three-wave Raman resonant process \cite{SSM2001}.
Multi-component vortex solitons sustained by the interaction of the
beams with different frequencies in both quadratic and cubic
materials are also well known, though under the most typical
conditions the finite radius vortex solitons  break into filaments
due to azimuthal instabilities
\cite{Desyatnikov,Buryak2002,SF1998,Torres1998a}.

While the  above mentioned experimental and theoretical research
of nonlinear vortex charge conversion has
focused on cases involving a small number of frequency
components, typically two or three, the efforts directed towards
short pulse generation have resulted in the development of  techniques
leading to the generation of dozens of coherent frequency side-bands,
by means of cascaded four-wave mixing in Raman active gases
\cite{SH2003,BCS2006}. The latter technique does not rely on  the
waveguide or cavity geometries to boost nonlinear interaction and is therefore suitable for the simultaneous
frequency and vortex charge conversion. This idea has been explored by our
group and we have recently demonstrated simultaneous
generation of frequency and vortex combs \cite{we_prl}
in a Raman medium excited off-resonance with the
two pump beams, when one of the two carries a unit vortex and the other is vortex free.
We have
derived the vortex conversion rules and demonstrated
that the simultaneous  frequency and vortex combs are shaped in the
form of the spatio-temporal helical beams \cite{we_prl}.
On the focusing side of the Raman resonance, the multi-component vortex solitons have been found.

The aim of this work is to  report
regular tracing of the multi-component vortex solitons in the parameter
space and  to study  their linear stability with respect to perturbations.
Our analysis shows  that the spectrally symmetric soliton solutions
centered around the vortex-free frequency component are typically
unstable, although the instability fully develops only after long propagation distances.
At the same time,  the asymmetric solitons, for example those where all the generated
components are the Stokes ones, have a broad  stability range.
Based on the results of the linear  stability analysis for 2 and 3 component
solitons, we  demonstrate the same general tendencies of the soliton dynamics  for
the case of many coupled side-bands.

\section{Model}
\label{sec:Model}
The dimensionless model describing the evolution of the  side-bands
in an off-resonantly excited Raman
medium is \cite{SH2003,we_prl}
\begin{equation}
\label{eqE}
%\hat{\mathcal{L}}_n E_n=Q^* E_{n-1}+ Q E_{n+1},
i\partial_{z} E_n-\frac12\Delta E_n=\beta_n E_n + Q^* E_{n-1}+ Q E_{n+1},
\end{equation}
where $n=-M+1,\dots,0,\dots, N$ $(M,N\ge 0)$,
%$\hat{\mathcal{L}}_n=i\partial_z - 0.5\Delta-\beta_n$,
 and $\Delta=\partial_x^2+\partial_y^2$.
$E_n$ are the dimensionless amplitudes of the sidebands, such that the total field
is given by
\begin{equation}
\label{etot}
E_{tot}=\sum_n E_n(x,y,z)e^{i\Omega_n t-iK_nz}\;,
\end{equation}
where $\Omega_n=(\omega_0+n\omega_{mod})/\omega_{mod}$,
$\omega_{mod}=\omega_1-\omega_0$ is the modulation frequency (i.e. the frequency
difference between the two driving fields). $N$ is the  number of the anti-Stokes components
and $M-1$  is the number of the Stokes components. Taking into account the $E_0$ field, we have $M+N$
interacting Raman side-bands.
The physical frequencies and wavenumbers are represented by the lower case letters
$\omega_n$ and $k_n$, whilst their dimensionless counterparts by the upper case:
 $\Omega_n$ and $K_n$.
The dimensionless time $t$ is measured in  units of  $1/\omega_{mod}$,
the propagation coordinate $z$ is in  units of $L$, and the
transverse coordinates $(x,y)$ are in  units of $\sqrt{Lc/\omega_0}$.
$K_n=(\omega_0+n\omega_{mod})L/c$ are the scaled free space wavenumbers.
Here,  $L=(\eta\hbar\omega_0{\cal N}|b|)^{-1}$ characterizes the coupling length over which power
is transferred between  neighboring side-bands in the absence of dispersion.
$\eta\approx 376$ is the free space impedance, $\cal N$ is the density of molecules and $b$ is a coefficient
characterizing the material dependent coupling  between the sidebands \cite{SH2003}. The
weak frequency dependence of $b$ is neglected for simplicity.

\begin{figure}
\includegraphics[width=0.45\textwidth]{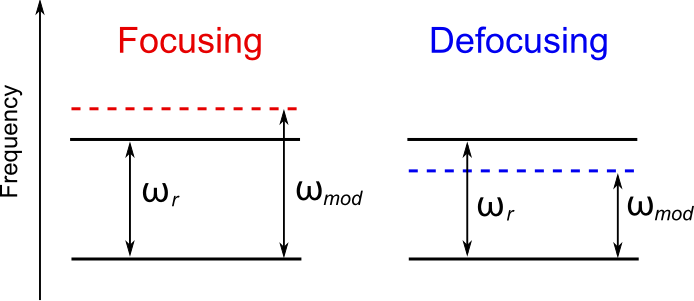}
\caption{(color online). Off-resonant excitations of the Raman transition creating either focusing
or defocusing nonlinearities. $\omega_r$ is the Raman frequency and $\omega_{mod}=\omega_1-\omega_0$, where
$\omega_1>\omega_0$ are the pump frequencies.}
\label{figSetup}
\end{figure}

$L$ varies from $1$ to a few mm for $D_2$ and $H_2$ gases \cite{SH2003},
so that one unit of $x$ corresponds to a few tens of microns.
$Q$ is the Raman coherence responsible for the coupling
between the side-bands.
Neglecting dissipation due to finite linewidth of atomic transition and finite dephasing time,
in the adiabatic approximation \cite{SH2003,we_prl,we_ol}
\begin{equation}
\label{stacQ}
Q(E_n)={sgn({\mu}) S \over 2\sqrt{\mu^2+|S|^2}},\quad S=\sum_nE_nE_{n+1}^*,
\end{equation}
where $\mu=|\omega_{mod}-\omega_r|/(|b|I_0)$
is the scaled modulus of the detuning of the modulation
frequency from the Raman frequency $\omega_r$. We also note, that
the above result is obtained under the assumption of equal Stark shifts of molecular levels, which
is the case for large detunings \cite{SH2003}.
While $|\mu|$ can always be fixed to unity by proper rescaling of the field amplitudes, its sign
controls the effective type of nonlinearity in Eqs.~(\ref{eqE}): positive (negative)
$\mu$ corresponds to the focusing (defocusing) nonlinearity \cite{we_prl, PUG+2008},
see Fig.~\ref{figSetup}. In what follows we consider the case of the focusing nonlinearity \cite{PUG+2008,WYS+2002},
($\mu=1$), which is known to support bright soliton solutions \cite{YWS2003,Yavuz2007a}.

$|Q|$ varies from $0$ to $1/2$ for $|S|/\mu$ varying from $0$ to $\infty$.
Therefore nonlinear interaction between harmonics is saturated
at high powers  or, equivalently, at small detunings $|\omega_{mod}-\omega_r|$.
$E_n\sqrt{I_0}$ are the dimensional amplitudes of the harmonics.
For $D_2$ and $H_2$ gases $\mu=1$ corresponds to  $I_0\sim 0.1$GW/cm$^2$,
provided  $|\omega_{mod}-\omega_r|\sim 1$GHz.
$\beta_n\equiv\beta(\omega_n)$ is the propagation constant
of the $n$th harmonic.

\section{Soliton solutions: General framework}
\begin{figure}
\includegraphics[width=0.23\textwidth]{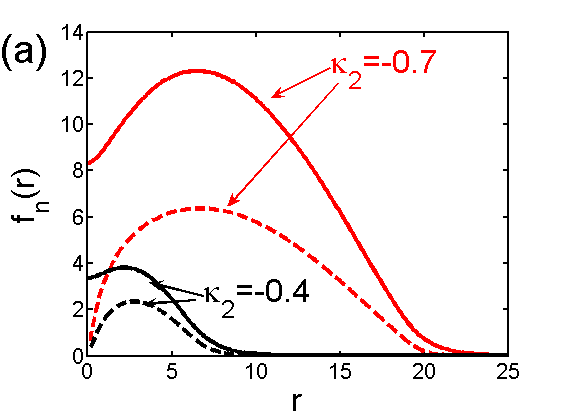}
\includegraphics[width=0.23\textwidth]{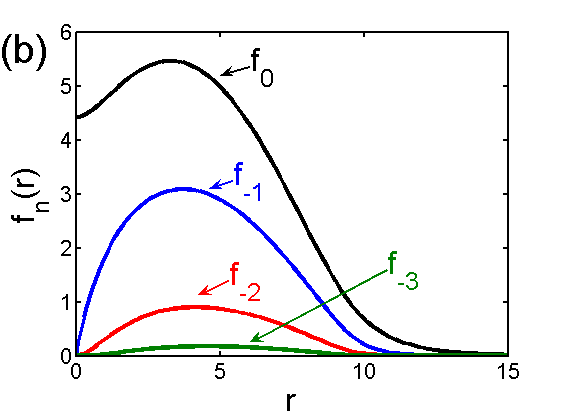}
\caption{(color online). Soliton radial profiles for the configuration with only Stokes components being excited ($N=0$),
$\beta_n=0.005n^2$ (which corresponds to normal dispersion), charge conversion step is unity: $l_n=n$.
(a) $2$ coupled fields ($M=2$), $\kappa_1=-0.25$, $\kappa_2=-0.4$ (black curves), $\kappa_2=-0.7$
(red/gray curves). Solid (dashed) curves correspond to $f_0$ ($f_{-1}$); (b) $5$ coupled fields ($M=5$),
$\kappa_1=-0.25$, $\kappa_2=-0.7$.
}
\label{figSolitonProfiles}
\end{figure}

In this and the next chapter we describe the general framework for finding
the stationary soliton solutions and studying their linear stability. Application
of these techniques to the cases of two and three components are described in detail
in Chapters V and VI.
The fact that equations (\ref{eqE}), (\ref{stacQ}) are invariant with respect to
$E_n\to E_n \exp(i\phi)$ and  $E_n\to E_n \exp(i n\psi)$, where $\phi$ and $\psi$
are arbitrary constants \cite{we_prl, we_ol}, implies the conservation of
the two integrals  $P=\sum_n I_n$ and
$R=\sum_n n I_n$, where $I_n=\int\int dxdy |E_n|^2$, and suggests the following ansatz for the soliton solutions:
\begin{equation}
\label{solitons}
E_n(x,y,z)=f_n(r)\exp\left[il_n\theta+i(\kappa_1+\kappa_2n)z\right].
\end{equation}
Here $r$ and $\theta$ are the polar radius and angle,
$l_n=l_0+n(l_1-l_0)$ is the vortex charge of the $n$th harmonic,
$\kappa_{1,2}$ are free parameters associated with the above symmetries.
The choice of $l_0$ and $l_1$ defines the step, $\Delta l=l_1-l_0$,
in which the vortex charge is changing  between the
adjacent side-bands. $f_n(r)$ are real functions obeying
\begin{eqnarray}
\nonumber
-\frac12\left[\frac{d^2f_n}{dr^2}+\frac{1}{r}\frac{df}{dr}-\frac{l_n^2}{r^2}f_n\right]
&=&\left[\kappa_1+\kappa_2 n +\beta_n \right]f_n\\
\label{eqF}
&+&q (f_{n-1}+ f_{n+1})\;.
\end{eqnarray}
where $q=Q(f_n)$. The boundary conditions  are \cite{SF1998}:
\begin{eqnarray}
\label{boundr0}
f_n(r)&\to& c_n^{(0)} r^{|l_n|} ,\quad r\to 0, \\
\label{boundr1}
f_n(r)&\to& c_n^{(\infty)} \frac{e^{-r\sqrt{-2(\kappa_1+\kappa_2 n + \beta_n)}}}{\sqrt{r}}, \quad r\to \infty\;.
\end{eqnarray}
where $c_n^{(0,\infty)}$ are real constants.
Eq. (6) naturally implies that the amplitude of a vortex carrying
component, $l_n\ne 0$, is zero at the phase singularity and
that the vortex free components, $l_n=0$, reach some constant value
at $r=0$. For the fields to decay to zero at $r\to \infty$,
one needs to select $\kappa_{1,2}$ to satisfy
\begin{equation}
\label{exist_condition}
\kappa_1+\kappa_2 n + \beta_n<0\;,
\end{equation}
simultaneously for all $n$. Without any loss of generality $\beta_0$
can always be set to zero by the  rotation of the common phase \cite{we_prl}.
Thus, fixing $\beta_0=0$ we find that
the above inequality for $n=0$ gives $\kappa_1<0$.
At the boundary points  of the above conditions
$f_n$ tends  to zero. Detuning
the $\kappa_{1,2}$ values  away from these boundaries into the range
where  $I_n$ is increasing eventually leads to the coherence tending
to its maximal value $q=1/2$.
Examples of the radial profiles of the vortex solitons are shown in Fig.~\ref{figSolitonProfiles}
for the asymmetric configuration with only Stokes components being excited ($N=0$).
When the propagation constants $\beta_n$ are symmetric around the central component: $\beta_n=\beta_{-n}$,
Eqs.~(\ref{eqF}) are invariant under the transformation $n\to-n$, $\kappa_2\to-\kappa_2$, $l_n\to-l_n$.
In this case solitons in the opposite
configuration, with only anti-Stokes components being excited ($M=1$), have exactly the same
structure as those shown in Fig.~\ref{figSolitonProfiles}.

In the vortex soliton case the $q=1/2$ limit is achieved not only
through the growing amplitudes, but also through the
expansion of the rings and flattening of their profiles.  The
soliton existence boundary corresponding to $q=1/2$ can be worked out neglecting the $x$ and $y$
dependence of $E_n$. Subsequently one can disregard the left-hand sides in
Eq. (\ref{eqF}), assume $q=1/2$ and work out constraints on the $\kappa_1$
and $\kappa_2$ values from the solvability conditions of the
resulting homogeneous equations: $[\kappa_1+\kappa_2n+\beta_n]f_n+(f_{n-1}+f_{n+1})/2=0$.
Examples of the existence domains
in the $(\kappa_{1},\kappa_2)$-plane can be seen in Figs. ~\ref{fig2fieldsExist}(a) and~\ref{figStab3}.

\section{Linear stability analysis: General framework}
Stability of the vortex solutions is of course an important problem,
since similar solutions in other models are known to exhibit strong
modulational instability along the rings
\cite{SF1998,Torres1998a,miold}. This instability can be suppressed
by nonlocal nonlinearities \cite{Briedis2005,Skupin2007}, and in
some cases  when the higher order nonlinearities (e.g. quintic) are
assumed to dominate  over the lower order ones (e.g. cubic), see,
e.g., \cite{Mihalache2002,Mihalache2003}. Our model is particularly
interesting because, as we will demonstrate below, it allows the
existence of a sufficiently broad parameter range, where stable
vortex solitons exist with the local type of nonlinearity derived
from the first principles. The latter is true since the nonlinearity
in Eq. (\ref{eqE}) is calculated from the  Schr\"odinger equation
for a Raman medium driven far from the resonance
\cite{SH2003,we_ol}.

In order to analyze the linear stability we add small perturbations $\epsilon_n$ to the
vortex solitons and substitute the following ansatz
\begin{equation}
\label{perturb}
E_n=\left[f_n(r)+\epsilon_n(r,\theta,z)\right]\exp\left[i(\kappa_1+\kappa_2 n)z+il_n\theta\right]\;.
\end{equation}
into Eqs. (\ref{eqE}). After linearization we find:
\begin{eqnarray}
\nonumber
&&i\partial_z \epsilon_n-\frac12
\left[
\partial^2_{rr}+\frac1r\partial_r+\frac{1}{r^2}
\left(
\partial^2_{\theta\theta}+i2l_n\partial_\theta-l_n^2
\right)
\right]\epsilon_n=\\
\nonumber
&&(\kappa_1+p_n)\epsilon_n+
q(\epsilon_{n-1}+\epsilon_{n+1})\\
\label{eqEpsilon}
&&
+\sum_m\left\{
A_{nm}\epsilon_m
+B_{nm}\epsilon_m^*
\right\}\;,
\end{eqnarray}
where
\begin{eqnarray}
\label{ankDEF}
A_{nm}&=&f_{n-1}M_m+
f_{n+1}P_k\;,\\
\label{bnkDEF}
B_{nm}&=&f_{n-1}P_m+
f_{n+1}M_m\;,\\
\label{Mdef}
M_m\equiv\frac{dq^*}{df_m}&=&\frac{\left(2\mu^2+s^2\right)f_{m-1}-s^2f_{m+1}}
{4\left(\mu^2+s^2\right)^{3/2}},\\
\label{Bdef}
P_m\equiv\frac{dq}{df_m}&=&\frac{\left(2\mu^2+s^2\right)f_{m+1}-s^2f_{m-1}}
{4\left(\mu^2+s^2\right)^{3/2}}\;,
\end{eqnarray}
$n,m=-M+1,\dots,0,\dots N$ and $s=S(f_n)$.

Expanding perturbations into azimutal harmonics \cite{SF1998}
\begin{eqnarray}
\nonumber
\epsilon_n(r,\theta,z)&=&\sum_{J\ge 0} \left\{
h^{+}_{n,J}(r,z)\exp(iJ\theta)\right.\\
\label{EpsExpand}
&+&\left.(h^{-}_{n,J}(r,z))^*\exp(-iJ\theta)
\right\}\;,
\end{eqnarray}
we assume $h^{\pm}_{n,J}(r,z)=g^{\pm}_{n,J}(r)
\exp(\lambda_{J} z)$ and derive the eigenvalue problem:
\begin{equation}
\label{eqG}
i\lambda_J \left[
\begin{array}{c}
\textbf{g}_J^+ \\
\textbf{g}_J^-
\end{array}
\right]
 =
\left[
\begin{array}{cc}
L^{+} & B \\
-B & -L^{-}
\end{array}
\right]
\left[
 \begin{array}{c}
\textbf{g}_J^+ \\
\textbf{g}_J^-
\end{array}
\right]
 \;,
\end{equation}
where
$\textbf{g}_{J}^{\pm}=\{g^{\pm}_{1-M,J}\;,\;g^{\pm}_{2-M,J}\;,\;\dots\;,\;g^{\pm}_{0,J}\;,\;\dots,
\;g^{\pm}_{N,J}\}^T$. $L^{\pm}$ and $B$ are the $(N+M)\times (N+M)$ matrix operators.
Elements of  $B$ are $B_{nm}$ and they are defined in Eq.~(\ref{bnkDEF}), and the elements of
$L^{\pm}$ are
\begin{eqnarray}
\nonumber
L^{\pm}_{nm}&=&\delta_{n,m}\left\{
\frac{1}{2r}\frac{d}{dr}\left(r\frac{d}{dr}\right)
- \frac{(J\pm l_n)^2}{2r^2}\right.\\
\label{Ldef}
&&
\left.
+\kappa_1+\kappa_2 n +\beta_n
\right\}\\
\nonumber
&&
+q\left(\delta_{n+1,m}+\delta_{n-1,m}\right)+A_{nm}\;,
\end{eqnarray}
where $\delta_{n,m}$ is the Kronecker symbol.
For a solution $f_n$ to be linearly unstable there must exist $\lambda_J$
with $Re(\lambda_J)>0$.
Boundary conditions for eigenstates $\textbf{g}^{\pm}_J$ are defined
in a similar way to the boundary conditions for $f_n$ (see Eqs.~(\ref{boundr0})-(\ref{boundr1})),
but with $l_n$ being replaced by $l_n\pm J$.
We solve the eigenvalue problem in Eq.~(\ref{eqG}) numerically, replacing
differential operators by the second-order finite differences.
Note that accurate stability analysis of the multi-component solutions is rather complicated.
Therefore we will reveal basic mechanisms of instabilities of coupled vortex
solitons by focusing  on two- and three-component configurations.
Then we will demonstrate by   numerical modeling of Eqs. (\ref{eqE}),
that the instability and stabilization mechanisms found in the simplest cases
can be seen in the multi-component dynamics.

\section{Two-component vortex solitons}
We start with the simplest configuration of
two side-bands, that is $n=0,1$ ($N=M=1$) in Eqs.~(\ref{eqE}).
This applies e.g. to the opposite circularly polarized driving fields $E_0$ and $E_1$,
when the cascaded generation of Stokes and anti-Stokes harmonics is forbidden due to angular
momentum selection rules \cite{YWS2003}. The propagation equations in this case are
\begin{eqnarray}
\label{eqE1_2fields}
\left(i\partial_{z} -\frac12\Delta -\beta_0\right)E_0&=&
\frac{|E_1|^2E_0}{2\sqrt{\mu^2+|E_0|^2|E_1|^2}}\;,\\
%i\partial_{z} E_0-\frac12\Delta E_0&=&\beta_0 E_0 + \frac{E_0 |E_1|^2}{2\sqrt{\mu^2+|E_0|^2|E_1|^2}}\;,\\
\label{eqE2_2fields}
\left(i\partial_{z} -\frac12\Delta -\beta_1\right)E_1&=&
\frac{|E_0|^2 E_1}{2\sqrt{\mu^2+|E_0|^2|E_1|^2}}\;.
%i\partial_{z} E_1-\frac12\Delta E_1&=&\beta_1 E_1 + \frac{E_1 |E_0|^2}{2\sqrt{\mu^2+|E_0|^2|E_1|^2}}\;.
\end{eqnarray}
Eqs. (\ref{eqE1_2fields}, \ref{eqE2_2fields}) explicitly express a known fact that
the fields interacting via the Raman
nonlinearity do not have nonlinear self-action. This property  does
not depend on the number of interacting components.

Bright (vortex free) spatial solitons in the two-component Raman model have been studied
in \cite{Yavuz2007a,YWS2003} and the associated self-focusing
effects have been observed in \cite{PUG+2008,WYS+2002}. Also, there are closely related
recent results on spatial solitons in Raman active liquids
\cite{Fanjoux2006,Fanjoux2008}. The papers \cite{time}
have analyzed the two-component temporal Raman solitons existing in the presence of group velocity dispersion,
i.e. when the transverse Laplacian  is replaced with the 2nd-order time derivative.
The above two-component model is also similar to that for the so-called holographic
solitons \cite{CCS+2002}.

\begin{figure}
\includegraphics[width=0.23\textwidth]{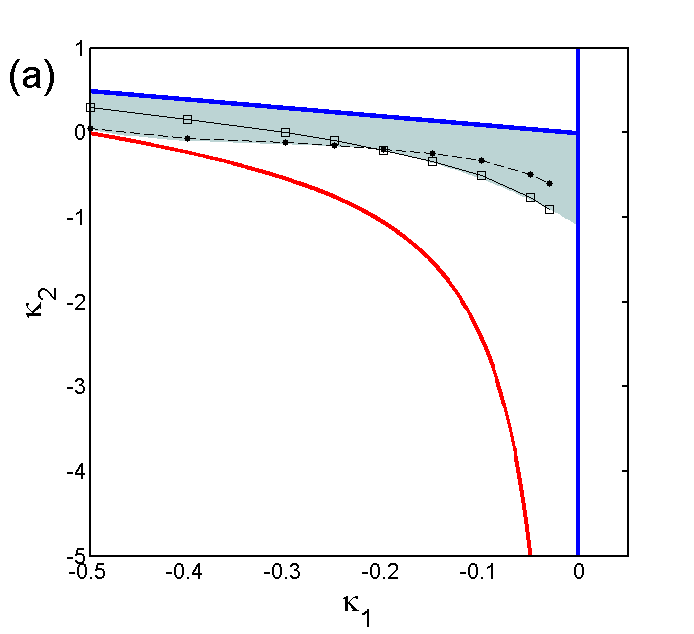}
\includegraphics[width=0.23\textwidth]{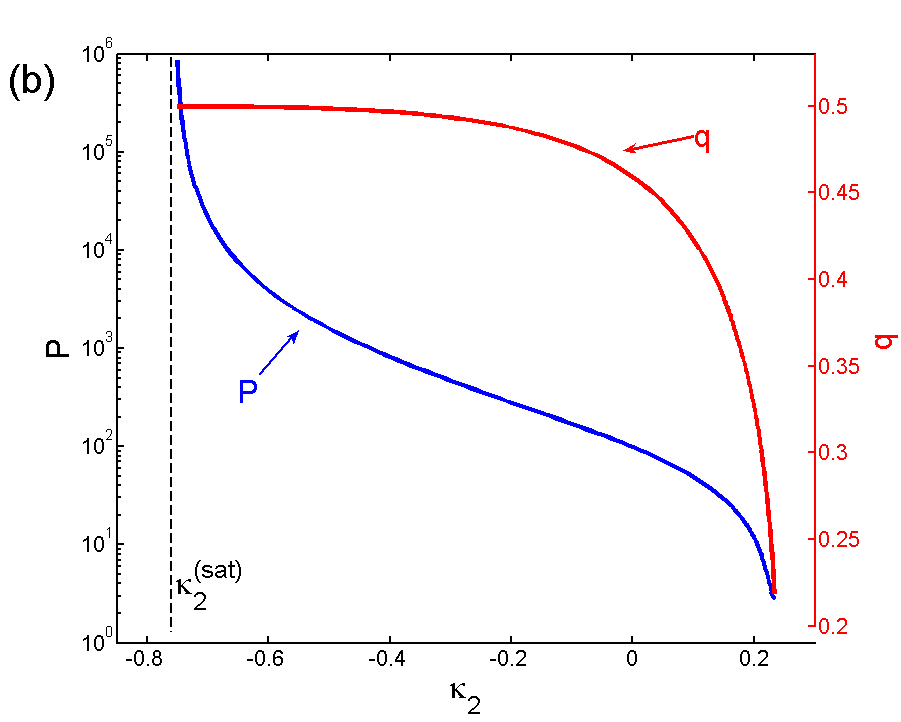}
\caption{(color online). (a) Region of existence of coupled vortex solitons for the case of two fields,
$\beta_1=0.005$.
Straight blue lines correspond to the boundaries of existence in Eq.~(\ref{exist_condition}),
red/gray curve - to the dispersion $\kappa_2^{(sat)}$ of high-intensity constant
amplitude waves, Eq.~(\ref{disp_high_intens}).
Shaded area indicates region of unstable solutions for the configuration $(l_0=0, l_1=1)$,
open squares and filled circles
correspond to numerically found instability thresholds for the $J=1$ and $J=2$
unstable perturbations, respectively (see main body text for details and
Figs.~\ref{fig2fieldsStab}, \ref{fig2fieldsJ1dynamo}, \ref{fig2fieldsJ2dynamo}).
(b) Soliton power $P$ and maximum value of the coherence $q$ versus $\kappa_2$
at fixed value of $\kappa_1=-0.25$.
Approaching the boundary $\kappa_2^{(sat)}(\kappa_1)$ $q$ is saturated at its maximum
value $|q|=0.5$, while the norm tends to infinity.
}
\label{fig2fieldsExist}
\end{figure}

The existence conditions for soliton solutions in Eqs.~(\ref{exist_condition}) are reduced to
the joint inequalities $\kappa_1<0$ and $\kappa_2<-\kappa_1-\beta_1$, which define a
semi-infinite region in $(\kappa_1,\kappa_2)$ bounded by the two rays, see Fig.~\ref{fig2fieldsExist}(a).
Another boundary is derived from the $q=1/2$ condition and is given by:
\begin{equation}
\label{disp_high_intens}
\kappa_2>\kappa_2^{(s)}=\frac{1}{4\kappa_1}-\kappa_1-\beta_1\;.
\end{equation}

\begin{figure}
\includegraphics[width=0.25\textwidth]{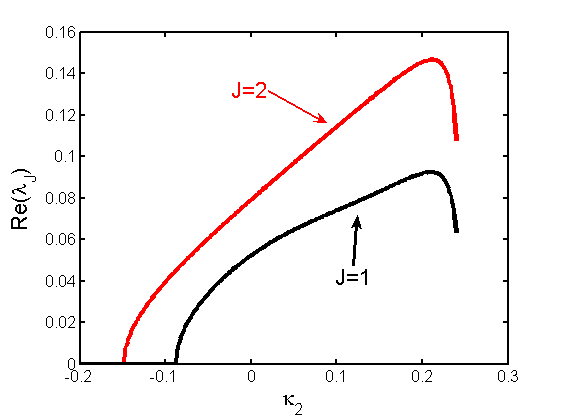}
\caption{(color online). Real part of eigenvalues corresponding to unstable perturbations versus $\kappa_2$ for
$\kappa_1=-0.25$.}
\label{fig2fieldsStab}
\end{figure}

Our linear stability analysis demonstrates that the soliton with $l_0=0$ and $l_1=1$
is unstable only inside the sufficiently narrow range of $\kappa_{1,2}$ corresponding
to the relatively small values of $q$, see Fig.~\ref{fig2fieldsExist}. As soon as $q$  increases and the
saturation effects  become important the solution becomes stable.
Note that the saturation of the
self-focusing nonlinearity does not stabilize the vortices in the models with the nonlinear self-action effects
\cite{SF1998}. It suggests that the absence of the self-action plays an important role in stabilization of
the vortex solitons.
In its instability range, the vortex soliton is  unstable
with two eigenvalues $\lambda_{J=1}$ and $\lambda_{J=2}$ having positive real
parts, see Fig.~\ref{fig2fieldsStab}.
Fixing $\kappa_1$ we numerically find the critical values of $\kappa_2$,
at which the two instabilities disappear, see circles and squares in Fig.~\ref{fig2fieldsExist}(a).
Selective numerical runs for the cases $l_0=0$,  $|l_1|\ge 2$ and $l_0=1$, $|l_1|\ge 1$
suggest that they are unstable with respect to azimuthal instabilities through large
parts of their existence domains.
\begin{figure}
\includegraphics[width=0.22\textwidth]{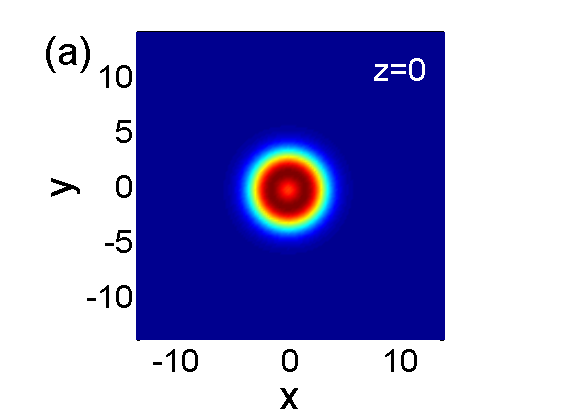}
\includegraphics[width=0.22\textwidth]{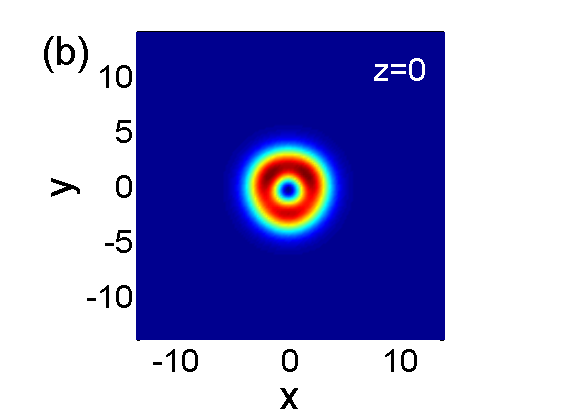}
\includegraphics[width=0.22\textwidth]{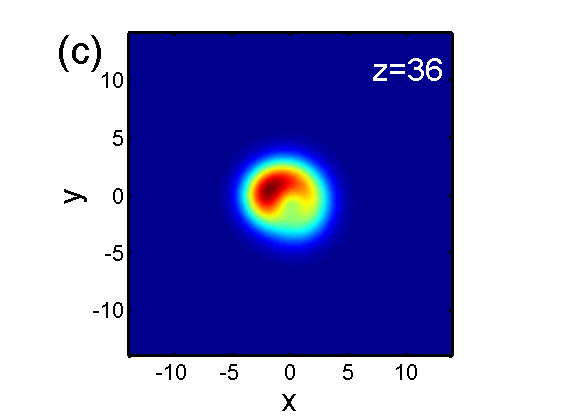}
\includegraphics[width=0.22\textwidth]{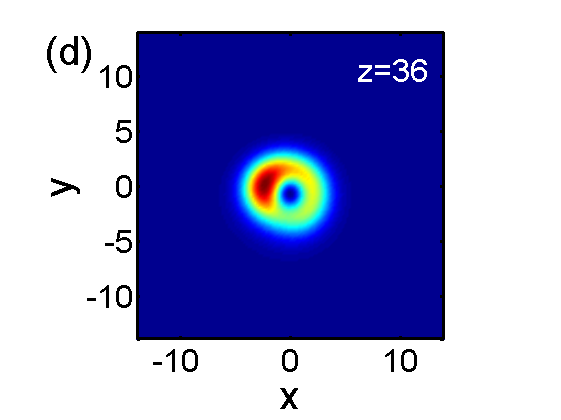}
\includegraphics[width=0.22\textwidth]{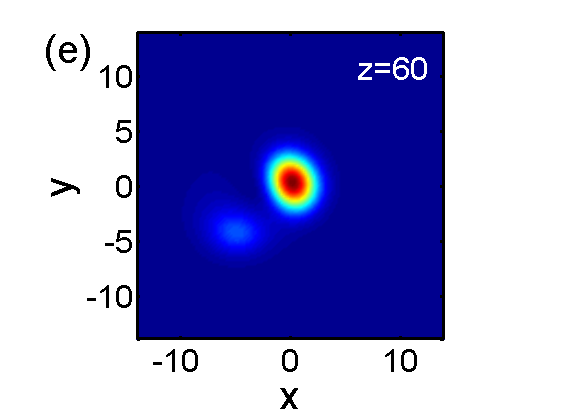}
\includegraphics[width=0.22\textwidth]{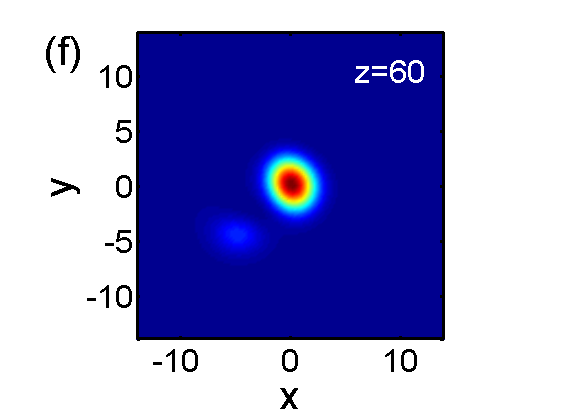}
\caption{(color online). Dynamics of the unstable soliton ($\kappa_1=-0.25$, $\kappa_2=0$) with
small perturbation along the $J=1$ unstable eigenvector. Cross-sections of
field intensities $|E_n|^2$ are plotted for the $0$th (left column) and
$1$st (right column) harmonics at different propagation distances $z$. The initial
soliton has a ``donut" shape, which is consistent with the vortex charge $l_1=1$ in the first harmonic.
As the instability evolves, it deforms the excitation towards usual bright spatial soliton
bearing no vortex charge. Note, that the overall orbital angular momentum is conserved and carried by rapidly diffracting
radiative waves.
}
\label{fig2fieldsJ1dynamo}
\end{figure}

\begin{figure}
\includegraphics[width=0.22\textwidth]{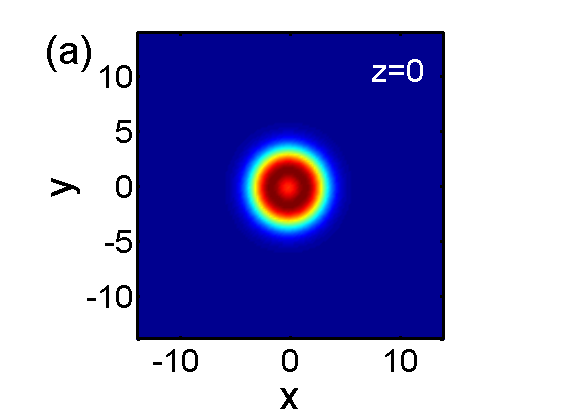}
\includegraphics[width=0.22\textwidth]{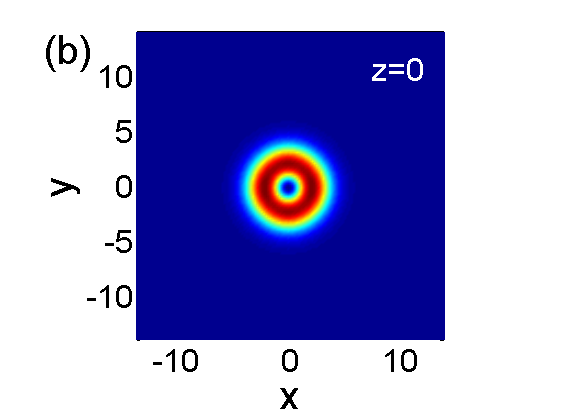}
\includegraphics[width=0.22\textwidth]{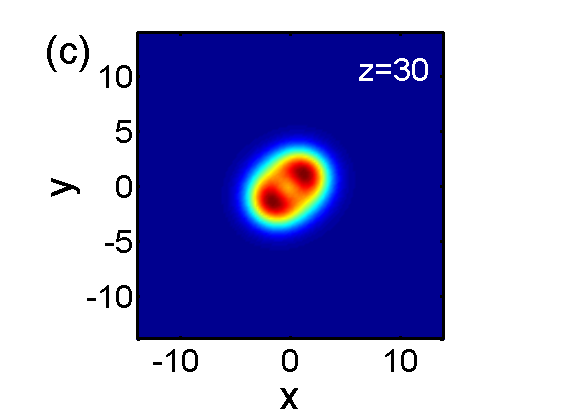}
\includegraphics[width=0.22\textwidth]{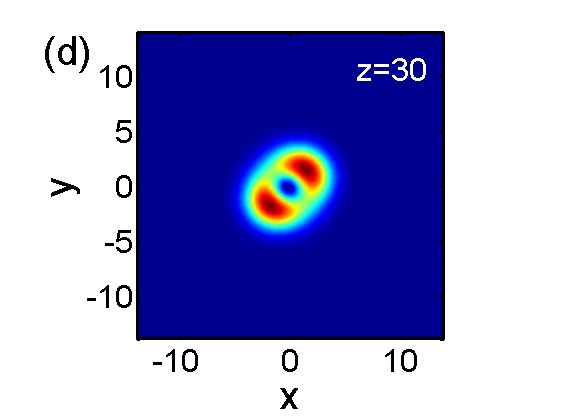}
\includegraphics[width=0.22\textwidth]{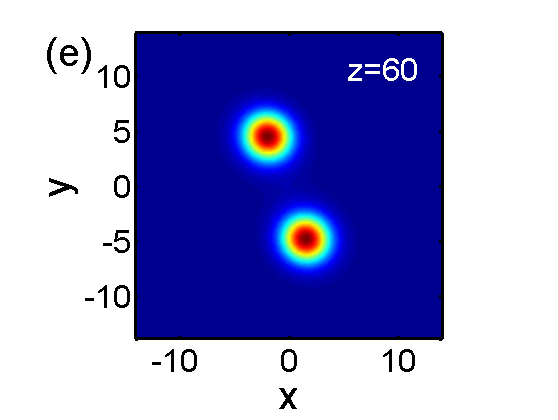}
\includegraphics[width=0.22\textwidth]{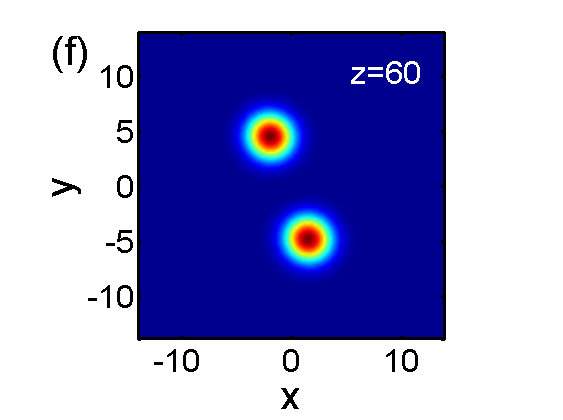}
\caption{(color online). The same as Fig.~\ref{fig2fieldsJ1dynamo} but with perturbation along $J=2$ unstable eigenvector.
As the instability evolves, the soliton is transformed into a pair of spatial solitons, which are then pulled apart
and moving in opposite directions.
}
\label{fig2fieldsJ2dynamo}
\end{figure}

To reveal the impact of instabilities on the soliton dynamics, we initialize Eqs.~(\ref{eqE})
with numerically found soliton solutions slightly perturbed along unstable eigenvectors
and perform dynamical simulations. Results are presented in Figs.~\ref{fig2fieldsJ1dynamo}
and \ref{fig2fieldsJ2dynamo} for the $J=1$ and $J=2$ unstable eigenvectors, respectively.
Both perturbations break the soliton symmetry and eventually lead to the formation
of a single or a pair of bright spatial solitons \cite{YWS2003,Yavuz2007a}.

\section{Three-component vortex solitons}
The addition of the third component makes the interaction between the Raman side-bands
phase-sensitive, and the choice of the vortex charges $l_n$ in any two fields
defines the charge of the remaining field via the phase-matching
conditions \cite{we_prl}.  Eqs. (\ref{eqE})
for the three component case with  $n=1-M,2-M,3-M$ are:
\begin{eqnarray}
\nonumber
&&\left(i\partial_{z} - \frac12\Delta -\beta_{1-M}\right)E_{1-M}=\\
\label{eqE1mk}
&&\qquad\frac{C}{2}\left[
|E_{2-M}|^2 E_{1-M} + E_{2-M}^2 E_{3-M}^*
\right]\;,\\
\nonumber
&&\left(i\partial_{z} - \frac12\Delta -\beta_{2-M}\right)E_{2-M}=\\
%i\partial_{z} E_{2-K}&-&\frac12\Delta E_{2-K}=\beta_{2-K} E_{2-K}\\
\nonumber
&&\qquad\frac{C}{2}\left[
\left(|E_{1-M}|^2+ |E_{3-M}|^2\right) E_{2-M} \right. \\
\label{eqE2mk}
&&\qquad \qquad \left.
+ 2 E_{3-M}E_{1-M} E_{2-M}^*
\right]\;,\\
\nonumber
&&\left(i\partial_{z} - \frac12\Delta -\beta_{3-M}\right)E_{3-M}=\\
%i\partial_{z} E_{3-K}&-&\frac12\Delta E_{3-K}=\beta_{3-K} E_{3-K}\\
\label{eqE3mk}
&&\qquad\frac{C}{2}\left[
|E_{2-M}|^2 E_{3-M} + E_{2-M}^2 E_{1-M}^*
\right]\;,
\end{eqnarray}
here $C^2=1/\{\mu^2+|E_{1-M}E_{2-M}^*+E_{2-M}E_{3-M}^*|^2\}$.
Fixing  $l_1-l_0=1$, we consider two  cases ($M=3$ and $M=2$):
asymmetric $(l_{-2}=-2,l_{-1}=-1,l_0=0)$ and
symmetric $(l_{-1}=-1,l_0=0,l_1=1)$.
The former corresponds to the often encountered case with negligible anti-Stokes
side-bands, and the latter implies that the first Stokes and first anti-Stokes
lines are excited.

\begin{figure}
\includegraphics[width=0.23\textwidth]{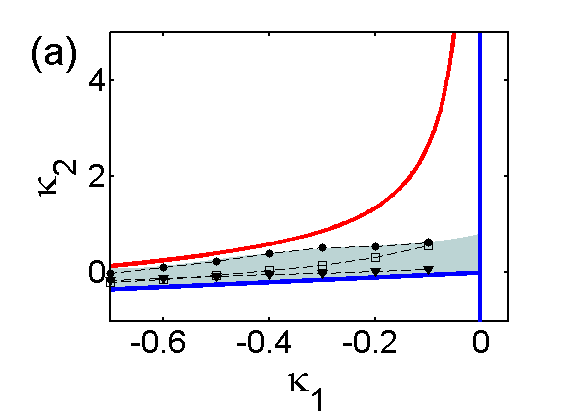}
\includegraphics[width=0.23\textwidth]{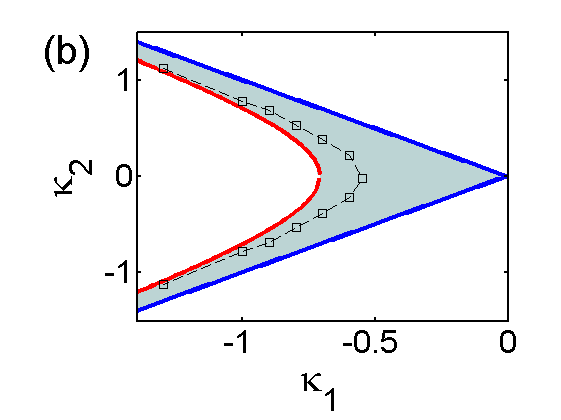}
\caption{(color online). Region of existence of coupled vortex solitons for the case of three fields:
(a) asymmetric configuration $(l_{-2}=-2,l_{-1}=-1,l_0=0)$, $\beta_{-1}=0.005$,
$\beta_{-2}=0.02$;
(b) symmetric configuration $(l_{-1}=-1,l_0=0,l_1=1)$, $\beta_{1}=\beta_{-1}=0.005$.
Open squares, filled circles and filled triangles
correspond to numerically found instability thresholds for $J=1$,
$J=2$ and $J=3$ unstable perturbations, respectively.
Shaded areas indicate regions of unstable solutions.}
\label{figStab3}
\end{figure}

The existence boundary for the asymmetric case given by the condition $q=1/2$
is now $\kappa_2<\kappa_2^{(s)}$, where
\begin{eqnarray}
\nonumber
\kappa_2^{(s)}=-\frac14\left\{
\frac{1}{\kappa_{1}}-3\kappa_{1}-2\beta_{-1}-\beta_{-2}\right.\\
\label{disp_high_intens_3fields_edge}
-  \left. \sqrt{
\left(\frac{1}{2\kappa_{1}}+\beta_{-2}-2\beta_{-1}-\kappa_{1}\right)^2+2-\frac{1}{4\kappa_{1}^2}
}
\right\}\;
\end{eqnarray}
In the symmetric case,
the  $q=1/2$ condition implies $\kappa_2^{(-)}<\kappa_2<\kappa_2^{(+)}$, where
\begin{equation}
\label{disp_high_intens_3fields_middle}
\kappa_2^{(\pm)}=\frac12\left\{\beta_-
\pm \sqrt{
\beta_-^2-\frac{\beta_+
\left(1-4\kappa_1^2\right)
}{\kappa_1}
-2+4\kappa_1^2
}
\right\}\;
\end{equation}
with $\beta_{\pm}=(\beta_{-1}\pm\beta_1)$.
Together with the  conditions in Eqs.~(\ref{exist_condition}),
the above constraints define the regions of the soliton existence, see Fig.~\ref{figStab3}.

Stability analysis demonstrates that, similar to the two-component
case with $l_0=0$ and $l_1=1$, the three-component solitons with
$l_{-2}=-2$, $l_{-1}=-1$, $l_0=0$ are stable inside a sufficiently
wide domain in the $(\kappa_1,\kappa_2)$ plane and, in particular,
in the proximity of the existence boundary given by $q=1/2$, i.e. in
the high saturation regime. Close to the lower boundary of the
existence domain given by $\kappa_2=(\kappa_1+\beta_{-2})/2$ there
are three types of instabilities with $J=1,2,3$, see
Fig.~\ref{figStab3}(a). 
We note that the solution with the
side-bands generated on the anti-Stokes side, i.e. the solution with
$l_0=0$, $l_1=1$, $l_2=2$, has the same stability properties as the
solution discussed above.
The symmetric case with $n=-1,0,+1$ is found
to be unstable with respect to the $J=1$ and $J=2$ instabilities,
with the former one persisting in the entire  existence domain, see
Fig.~\ref{figStab3}(b). 

\section{Multi-component vortex solitons and  spatio-temporal helical beams}
The above results show that if the vortex soliton contains a vortex
free component, for example, at $n=0$, and vortex carrying
side-bands either only on the Stokes or only on the anti-Stokes
sides, it can be stable within a broad range of parameters
$\kappa_{1,2}$ ensuring that the saturation effects are sufficiently
strong. Since in the frequency comb generation experiments with the
off-resonant Raman  gases the total number of excited harmonics can
go to a few dozen \cite{SH2003,BCS2006}, an important question to be
addressed is whether the above stated principles of the vortex
soliton stabilization can be extended onto multi-component cases. To
address this problem we use numerical integration of
Eqs.~(\ref{eqE}) with $11$ coupled side-bands, initialized with the
three-component vortex solitons described in the previous section.
We consider two cases: (i) asymmetric case where excitation of the
anti-Stokes lines is suppressed, and (ii) symmetric case with
excitation of Stokes and anti-Stokes lines being equally probable.
In both cases we number the harmonics in a way that $n=0$
corresponds to the vortex-free component. Thus we take $M=11, N=0$
and $M=6, N=5$  in Eqs.~(\ref{eqE}) for the asymmetric and symmetric
cases, respectively.

We monitor the evolution of the fields by plotting the total field intensity $I_{tot}=|E_{tot}|^2$ with $E_{tot}$
defined in Eq.~(\ref{etot}). It has been demonstrated in \cite{we_prl} that simultaneous
frequency and vortex combs lead to the  helical structure
of the total field intensity $I_{tot}$, both in $(x,y,t)$ and $(x,y,z)$ subspaces.
For the case of initial conditions where all the fields apart from
the three pumps $k-1$, $k$ and $k+1$ are initially zero, the $I_{tot}$ can be crudely  approximated with
[see Appendix for details]:
\begin{equation}
I_{tot}(x,y,z,t)\approx \left|
f_{k}^{(0)}+
f_{k-1}^{(0)}e^{-i\phi}
+f_{k+1}^{(0)}e^{i\phi}
\right|^2\;,
\label{Itot}
\end{equation}
where $\phi=t+\Delta l \theta - Kz$, $\Delta l = l_{k+1}-l_k$ is the vortex charge  step between the neighboring side-bands
and $K= \omega_{mod} L/c-\kappa_2$. For any fixed $t$ and $z$ the total intensity distribution
in the transverse plane is modulated in $\theta$
with the period defined by $\Delta l$, and it rotates in both $t$ and $z$, forming a spatio-temporal helix.

\begin{figure}
\includegraphics[width=0.23\textwidth]{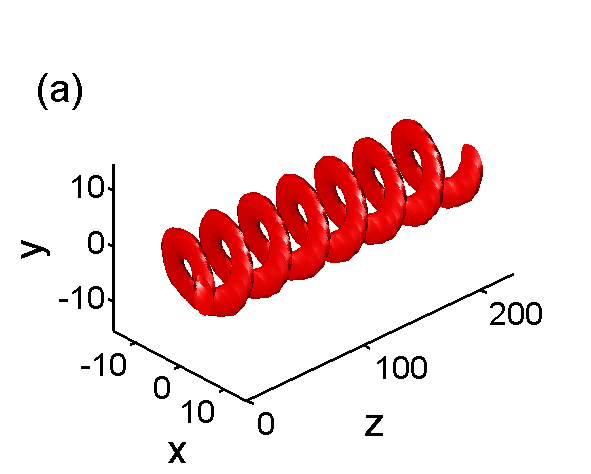}
\includegraphics[width=0.23\textwidth]{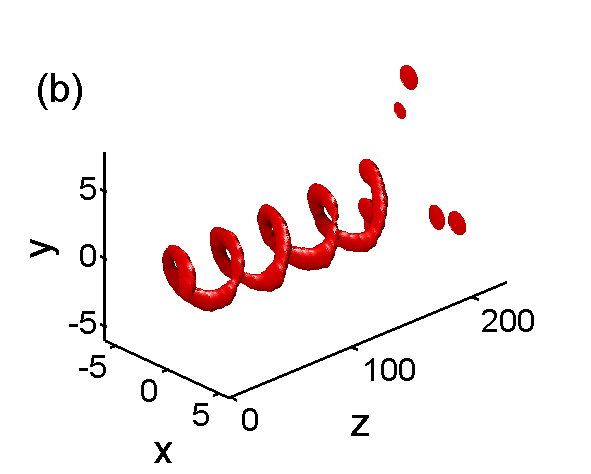}
\includegraphics[width=0.23\textwidth]{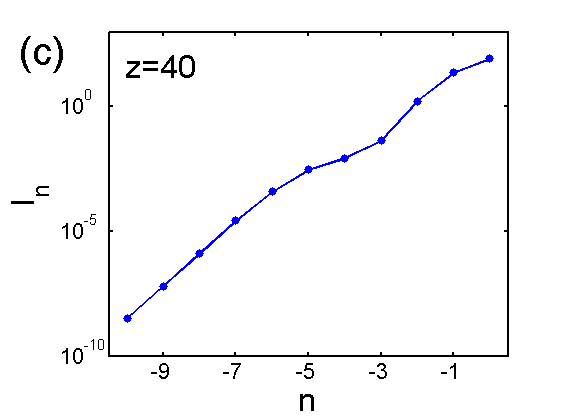}
\includegraphics[width=0.23\textwidth]{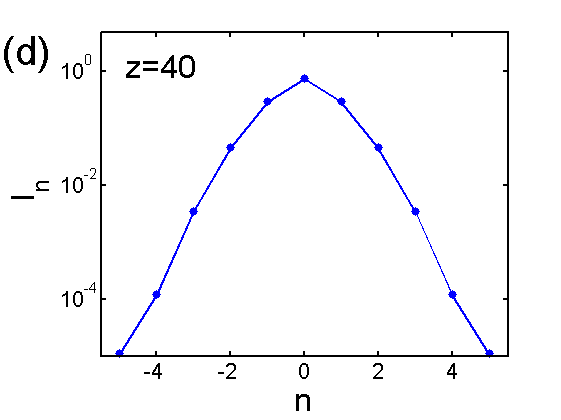}
\caption{(color online).
Dynamics of $11$ coupled fields initially excited with $3$-component solitons. (a) $I_{tot}(x,y,z,t=0)$
isointensity $(x,y,z)$-plot at 80\% at maximum for asymmetric configuration. Fields $n=-2,-1,0$
are initialized with the soliton, $\kappa_1=-0.25$, $\kappa_2=0.7$
[stable for $3$-component configuration, cf. Fig.~\ref{figStab3}(a)];
(b) The same as (a) but for symmetric configuration, fields $n=-1,0,1$ are excited with the soliton,
$\kappa_1=-0.25$, $\kappa_2=0$
[unstable for $3$-component configuration, cf. Fig.~\ref{figStab3}(b)]. Isointensity plot is at 60\% at maximum;
(c) and (d) intensity distribution
over harmonics after propagation distance $z=40$ for the cases in (a) and (b), respectively.
}
\label{figHelix}
\end{figure}

\begin{figure}
\includegraphics[width=0.23\textwidth]{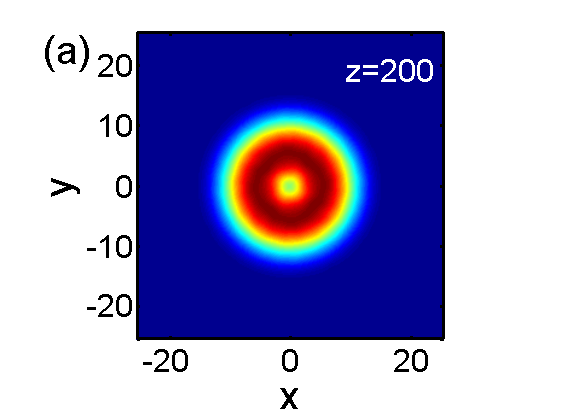}
\includegraphics[width=0.23\textwidth]{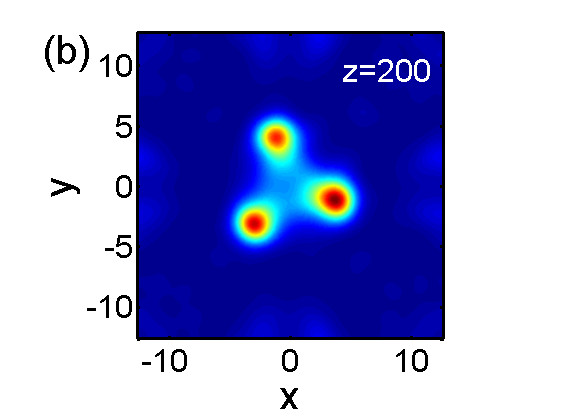}
\caption{(color online). Intensity distribution in the transverse plane of the $0$th harmonic after propagation distance $z=200$ for the asymmetric (a) and symmetric (b) configurations in Fig.~\ref{figHelix}.
Impact of the $J=3$ instability in the case of symmetric configuration
is clearly observed.
}
\label{figMultiDynamoFin}
\end{figure}

\begin{figure}
\includegraphics[width=0.4\textwidth]{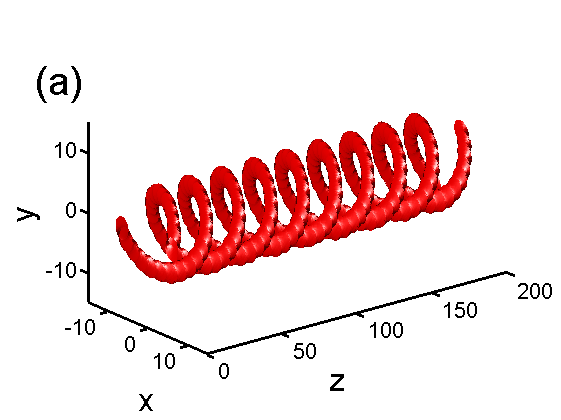}
\includegraphics[width=0.4\textwidth]{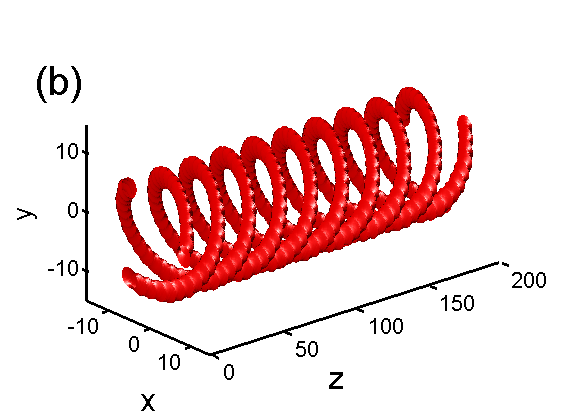}
\caption{(color online).
Stable double- (a) and triple-strand (b) helical beams formed by solitons with
$\Delta l=2$ and $\Delta l=3$, respectively. All the parameters are the same as in Fig.~\ref{figHelix}(a).
}
\label{figHiHelix}
\end{figure}

$\Delta l=1$ corresponds to  the single-strand helical  structure of $I_{tot}$, see
Fig.~\ref{figHelix} (a) and (b).
Fig. \ref{figHelix} (a) shows the long distance evolution of the helix in the
case of the asymmetric excitation,
with all the side-bands generated on the Stokes side, see the corresponding
spectrum in Fig. \ref{figHelix} (c).
The resulting helix in this case  keeps its structure
fixed over considerable propagation lengths. A similar numerical experiment for the
symmetric excitation results in the helical soliton, which breaks up into filaments after the same
propagation distance, cf. Figs. 8(a) and (b). Note, however, that the total length
in the simulations shown in Fig. \ref{figHelix} corresponds to a physical distance of order $20$cm,
which implies that one can speak about a quasi-stable propagation of the helix even  in the
case of the symmetric excitation of the Stokes and anti-Stokes side-bands.
The $z$-period of the helix, $2\pi/K$,
is not a  parameter of our numerical model, and it is only important when we are calculating $E_{tot}$.
Physically realistic values of the adimensional period are of the order of $1$ (for a typical modulation
frequency $\omega_{mod}$ of the order of $100$ GHz \cite{SH2003}), which makes the helical structure
contain several hundred periods over the distance of $180$ adimensional units
required to see the instability. Therefore, to make the structure of the helices and the break-up process more
obvious to the reader, we have fixed $K\approx 0.1$, when we have been producing
the images of the helices in Figs.~\ref{figHelix} and \ref{figHiHelix}.

Providing the asymmetric excitation conditions and changing  $\Delta
l$ to 2 and 3, we have also observed the formation of the stable
double- and triple-strand helices,  see Fig.~\ref{figHiHelix}.  Note
that the formation of similar multiple-strand helices has been
reported in \cite{FLP+2007}, as a result of the linear superposition
of the higher order Laguerre-Gauss modes. The helical soliton beams
reported here are qualitatively different from the so-called
spiraling solitons or rotating soliton clusters \cite{spir1, spir2, spir3, spir4, spir5}, which
sustain their rotation due to the interaction between the individual
beams accompanied by the conservation of the angular momentum. In
our case the helical evolution does not require the presence of more
than one intensity lobe, as shown in Fig.~\ref{figHelix}, and
originates from the interaction of multiple frequency harmonics
carrying progressively growing vortex charges. Most close known to
us analogue of the spatio-temporal helices studied above have been
reported in the context of the sine-Gordon equation and can be
observed in a chain of coupled pendulums \cite{book}.

\section{Summary}
In this work we have reported existence  conditions and have carried
out linear stability analysis of the two and three component vortex
solitons in an off-resonant Raman medium. We have found that, in the
case where the vortex carrying Raman side bands are located either
only on the Stokes or only on the anti-Stokes side of the vortex
free component, the vortex solitons have a significant stability
domain, corresponding to parameter values ensuring sufficient levels
of nonlinearity saturation. We have also demonstrated  that the same
stabilization mechanisms work in the case of many side-band, leading
to the excitation of stable helical beams with single-, double-, and
triple-strand topologies.

\section*{Appendix}
An approximate expression for the $z-$evolution of the simultaneous
frequency and vortex combs, excited with finite number of the
side-bands, can be found if one neglects diffraction and dispersion.
We replace Eq.~(\ref{solitons}) with $E_n(x,y,z)\approx
f_n(z)e^{il_n\theta}$ and use the fact that under these
approximations
\begin{equation}
\label{a1}
i\frac{\partial f_n}{\partial z}=q \left(f_{n+1}- f_{n-1}\right)\;.
\end{equation}
A solution to an initial value problem for  Eqs. (\ref{a1}) can be expressed using the
Bessel functions $J_n(z)$.
For an initial excitation with $N_0$ adjacent side-bands: $f_n^{(0)}\ne 0$ for $n=k,k+1,...,k+N_0-1$,
the resulting solution is given by
\begin{equation}
\label{a2}
f_n(z)=\sum_{j=k}^{k+N_0-1}f_{j}^{(0)}e^{-i\pi(n-j)/2}J_{n-j}(2q_0z)\;,\\
\end{equation}
where $q_0=q(z=0)$. The simplest case $N_0=2$ has been considered in \cite{we_prl,SH2003}.
Using the orthogonality of the Bessel functions:
$\sum_n J_{n+p}J_{n+q}=\delta_{p,q}$, it is easy to show that $q(z)\equiv q_0$
and thus  Eq.~(\ref{a2}) satisfies Eqs.~(\ref{a1}) for all $z$.
Substituting the  solution (\ref{a2}) into Eq. (\ref{etot}), we find
the approximate  expression for the total field:
\begin{eqnarray}
\nonumber
E_{tot}\approx \exp(i\phi_0)\sum_n\left\{
\exp\left[i n \phi \right]
\times
\right.\\
\label{a3}
\left.
\sum_{j=k}^{k+N_0-1}f_{j}^{(0)}e^{-i\pi(n-j)/2}J_{n-j}(2q_0z)
\right\}\;,
\end{eqnarray}
where $\phi_0=l_0\theta+\omega_0 t/\omega_{mod}-K_0z$,
$\phi=\Delta l \theta + t - K z$,
$K_0=\omega_0L/c-\kappa_1$,
$K=\omega_{mod}L/c-\kappa_2$. Using a known identity,
$\sum_n J_n(x)\exp(in\alpha)=\exp[ix\sin(\alpha)]$, we derive
\begin{equation}
\label{a4}
E_{tot}\approx \exp[i\phi_0+i 2q_0z\cos(\phi)]
\sum_{j=k}^{k+N_0-1}f_{j}^{(0)}\exp[ij\phi],
\end{equation}
which is the expression used in Eq. (\ref{Itot}).

\end{document}